\documentclass[aps,pre,reprint,floatfix,groupedaddress,amsfonts,amsmath,amssymb,citeautoscript]{revtex4-2}
\usepackage[pdftex]{graphicx}
\usepackage{color}
\usepackage{hyperref}
\usepackage{algorithm}
\usepackage[rightComments=false,noEnd=false,indLines=true]{algpseudocodex}

\begin{document}

\title{Algorithms for generating planar networks simulating hierarchical patterns of cracks formed during film drying}

\author{Yuri Yu. Tarasevich}
\email[Corresponding author: ]{tarasevich@asu-edu.ru}
\affiliation{Laboratory of Mathematical Modeling, Astrakhan Tatishchev State University, Astrakhan, Russia}

\author{Andrei V. Eserkepov}
\email{dantealigjery49@gmail.com}
\affiliation{Laboratory of Mathematical Modeling, Astrakhan Tatishchev State University, Astrakhan, Russia}

\author{Andrei S. Burmistrov}
\email{ksairen10@mail.ru}
\affiliation{Laboratory of Mathematical Modeling, Astrakhan Tatishchev State University, Astrakhan, Russia}

\date{\today}

\begin{abstract}
Hierarchical crack patterns that arise during the drying of thin films of colloidal dispersions or polymer solutions on a solid substrate are of interest both from a fundamental standpoint and in the context of the creation of transparent electrodes for optoelectronics. This paper analyzes the morphology of such patterns based on image processing of real-world samples. Three approaches are used to generate artificial hierarchical networks: random homogeneous tessellation, recursive Voronoi tessellation, and a crack growth simulation model, each modified to reproduce the hierarchical structure. A comparison was made of the geometric characteristics (distribution of crack angles, edge lengths, cell areas, and circularity coefficient) and topological properties (distribution of the number of cell sides) of real and simulated networks. It was shown that the simulation model best reproduces the key features of real cracks, including the characteristic right angles of their connections.
\end{abstract}

\maketitle

\section{Introduction\label{sec:intro}}
\begin{figure*}
  \centering
  \includegraphics[width=\textwidth]{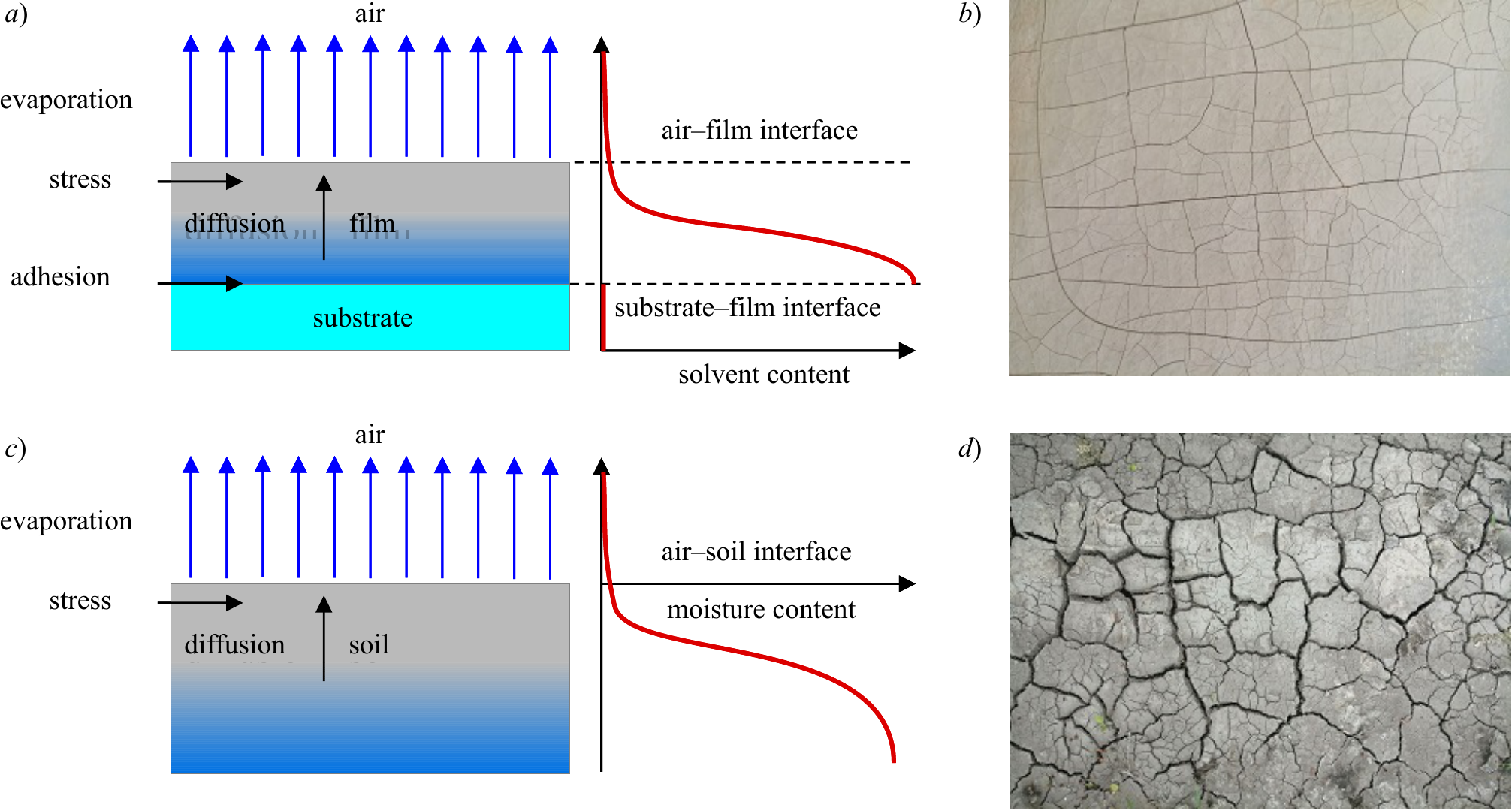}
  \caption{a) When a thin film of a colloid or a polymer desiccates on a horizontal solid substrate, a gradient of solvent concentration occurs inside this film due to evaporation of solvent from the free surface of the film along with diffusion of solvent from the bottom of the film to its top. Solvent loss leads to a shrinkage of the film and emergence of mechanical stress inside the film. b) Example of a crack pattern (craquelure) in painting. c) When soil desiccates, a gradient of moisture concentration occurs inside soil due to evaporation of water from the free surface of the soil along with diffusion of moisture from the depth to the surface. Moisture loss leads to a shrinkage of the soil and emergence of mechanical stress inside the soil. d) Example of desiccation crack pattern in soil.}\label{fig:film}
\end{figure*}

Transparent electrodes based on crack-templated metallic networks are widely used for next-generation optoelectronics~\cite{Cama2025a}. Understanding the morphology of crack patterns is important for production of transparent electrodes with required physical properties.
Patterns of cracks can emerge in thin films of colloidal dispersions and polymer solutions as they undergo desiccation on solid substrates~\cite{Gupta2025,Cama2025}. Evaporation of the solvent induces the film to shrink laterally, generating internal stresses. Cracks in the film appear when this mechanical stress exceeds a critical value that depends upon the interfacial energy and tensile properties of the material~\cite{Akiba2017} (Fig.~\ref{fig:film}a). Although crack patterns of soils and films appear similar and their fundamental formation mechanisms are the same, there are key differences, namely, absence of a substrate leads to a continuous vertical gradient of the moisture concentration in the case of a soil (Fig.~\ref{fig:film}c). In contrast, in films a sharp discontinuity in solvent concentration occurs at the film--substrate interface (Fig.~\ref{fig:film}a).

\begin{figure}[!tb]
  \centering
  \includegraphics[width=\columnwidth]{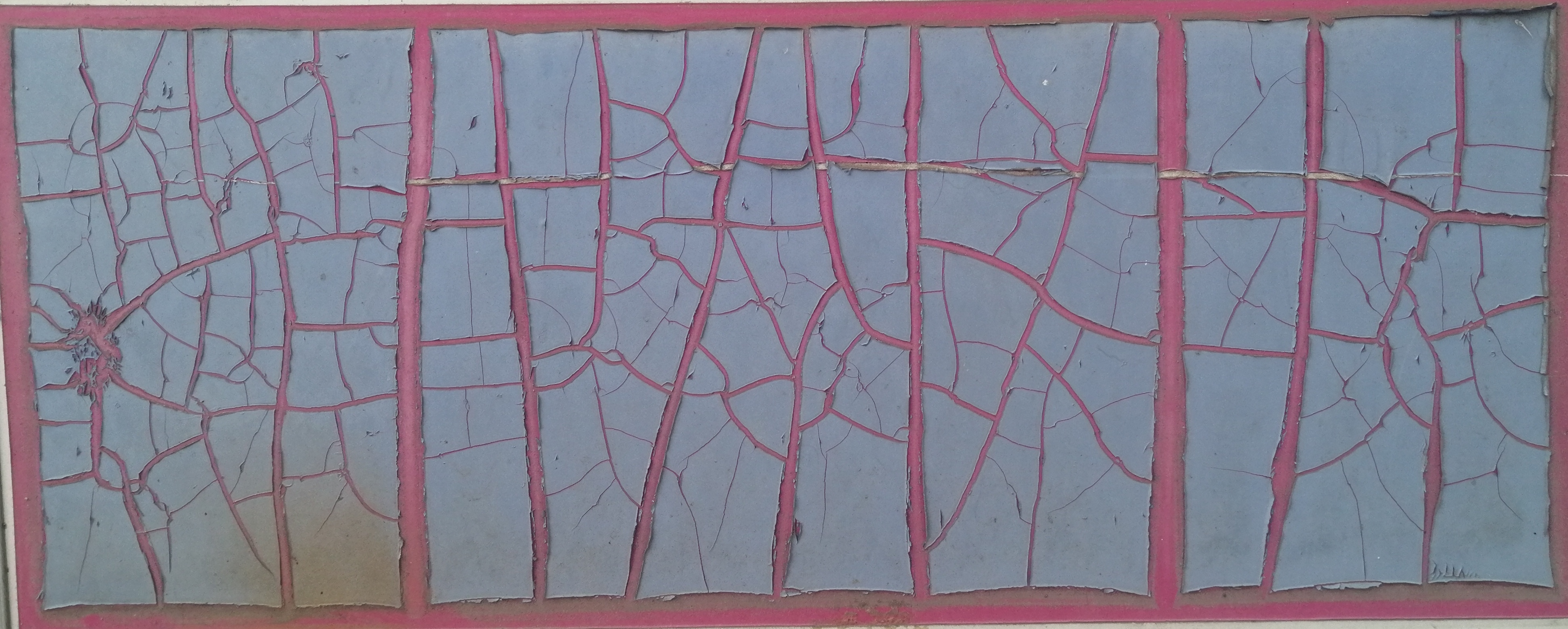}
  \caption{Clear hierarchical crack pattern on a painted wall.}\label{fig:wall}
\end{figure}

In a number of situations the pattern of cracks that is formed exhibits a clear hierarchical structure~\cite{Bohn2005,Bohn2005a,Cohen2009,Perna2011,Tang2011,Kumar2021,Voronin2022} (see also Fig.~\ref{fig:wall}). Based upon observations reported in Refs.~\onlinecite{Lazarus2011,Pauchard2020,Pauchard2020a,Guan2025}, such hierarchical crack patterns can be observed when the film thickness, $h$, exceeds a critical value $h_f$ that depends upon the moduli and surface energies of the film and interaction with the substrate. When the film thickness is less than another critical thickness, $h_c$, the film remains entirely free of cracks during the desiccation process~\cite{Guan2025}. For films of intermediate thicknesses, i.e., when $h_c<h<h_f$, stellar cracks occur due to nucleation. These cracks may be either isolated, or form a network without any visible hierarchy. As the film thickness increases, the patterns formed by the cracks evolve from isolated stellar cracks through disconnected sinuous cracks to partially connected crack networks, and, finally, complete connected network. In addition, spiral cracks can arise when the film thickness exceeds the critical value $h_s$~\cite{Lazarus2011}.

It is reasonable to expect that the evolution of crack patterns with the film thickness is due to the interplay of two processes, namely, the gelation of the solute (or suspended colloidal particles) and the evaporation of the solvent, with distinct characteristic timescales: the gelation time, $t_g$, and the evaporation time, $t_e$~\cite{Pauchard1999}. A number of the factors that affect crack patterns are enumerated in Ref.~\onlinecite{Gupta2025}.

Within a classification proposed in Ref.~\onlinecite{Gray1976}, when two cracks connect at an angle close to $90^\circ$, such crack junctions are referred to as being of the $T$-type. In addition, nuclei (defects) may form from which cracks begin to grow in the form of a three-pointed star with angles between cracks of approximately $120^\circ$. The classification proposed in Ref.~\onlinecite{Gray1976} describes such crack junctions as belonging to the $Y$-type. According to~\cite{Gray1976}, $X$-type crack junctions are common (four-pointed stars) in natural crack patterns, but higher-order junctions are generally absent. It can be assumed that $X$-type crack junctions are either (i) degenerate cases of two $T$-type cracks, when the edge between consecutive $T$-junctions is very short, or (ii) four-pointed stars extending from a shared nucleus.

The resulting desiccation crack patterns on solid substrate can be filled with a conductive material (Ag, Cu, Ni, etc.), which leads to the creation of a random conductive network (transparent electrode, transparent conductive film)~\cite{Voronin2023}. On the one hand, knowledge of the morphology of the hierarchical crack patterns is necessary, in particular, for calculating electrical conductivity, since cracks of different orders have different thicknesses~\cite{Kumar2021,Voronin2023}: the higher the order, the thinner the crack. On the other hand, without such knowledge, predictive modeling of the properties of hierarchical crack patterns is impossible, since it requires generating plausible crack networks on a computer.

A crack obviously tends to propagate along a maximal stress. The stress is expected to increase towards discontinuities (both domain boundaries and cracks). However, this stress field is not steady; on the contrary, the stress field changes as a crack propagates.

As a matter of fact, there are two competitive effects, namely, a stress field and an inertia of crack propagation. If a crack propagates very slowly, an intersection of cracks is expected to be strictly orthogonal, i.e., almost all crack junctions are of $T$-type (elastoplastic media; hierarchical crack pattern).  If the propagation is very fast, the crack has to go strictly along its original direction (brittle media; no bent cracks; almost all cracks are strictly linear; a tessellation is appropriate). In other words, the faster the crack propagation, the narrower the influencing area of boundaries; the slower the propagation, the broader the influencing area. This interpretation is based on the assumption that almost all cracks start from linear discontinuities (outer boundaries of the domain or already existing cracks). The interpretation omits the case where almost all cracks occur due to nucleation (three-arm stars start from point defects, i.e., point discontinuities) then merge.

The wide variety of observed crack patterns requires a similar variety of algorithms for their accurate modeling. The algorithms for modeling crack patterns formed in different materials can be divided into two groups, namely, approaches based on consideration of physical mechanisms and phenomenological approaches. The former uses either stress fields~\cite{Iben2009,Pons2010,Zhang2018,Liu2024} or discrete spring network models~\cite{Kitsunezaki1999,Vogel2005,Sadhukhan2007,Richardi2010,Sadhukhan2011,Khatun2012,Sadhukhan2019,Haque2023,Noguchi2024}. The latter uses STIT (stable with respect to iterations of tessellations)~\cite{Nagel2008,Leon2023}, Voronoi tessellation~\cite{Roy2022,Fanfoni2025}, Gilbert tessellation~\cite{Roy2022}, an evolution model for polygonal tessellations~\cite{Balint2023}, iterative cell division algorithm~\cite{Roy2022}, and simulations~\cite{Hafver2014,Yang2025b}. The appearance of the patterns obtained as a result of tessellations is very different from real-world hierarchical crack patterns, namely, (i)~the cracks are strictly linear, whereas in real-world systems they are curved and sometimes wavy; (ii)~there is an absence of $T$-junctions, which are typical in real-world hierarchical crack patterns, while sharp angles between cracks are abundant; (iii)~the cells between cracks vary greatly in size, which is not typical for real-world hierarchical crack patterns. The same behavior can also be observed in the case of model~\cite{Hafver2014}. The patterns that result from the spring model depend significantly on the geometry of the original spring network. Simulation~\cite{Yang2025b} generates crack patterns whose appearance is more typical of soils than of colloidal films and needs tuning of numerous parameters. Since the spatial variations in the properties of real films are unknown, the use of accurate and rigorous methods based on stress field calculations appears pointless. At least, it appears that these methods are primarily used for computer visualization rather than physical research.
In Ref.~\onlinecite{Roy2022}, the properties of macroscopic real-world crack patterns in various materials were compared with those of artificial computer-generated networks (planar graphs) that model these patterns; to generate these artificial analogues, three models were used, namely, Gilbert tessellation, Voronoi tessellation, and cell division. To characterize the samples, the authors used distributions of the isoperimetric ratio, the angular defect, the crack length, the number of vertices per polygon, and the polygon area.

The present work is devoted to two interconnected tasks. On the one hand, we analyze published images of microscopic hierarchical crack patterns in order to obtain clear fingerprints which can potentially be used for automatic classification of the hierarchical crack patterns.  On the other hand, we perform simulation of spatial-temporal evolution of hierarchical crack patterns in thin desiccating films. Since prepatterned surfaces significantly affect the crack pattern~\cite{Nandakishore2016,Lee2022} (e.g., craquelures in paintings on canvas or woody panels), we assume the substrate to be isotropic, homogeneous, and sufficiently smooth with strong adhesion to the film. Boundary conditions can provoke particular crack patterns (e.g., crack patterns of mud or clay in circular containers~\cite{Khatun2015} and of precipitates of desiccated drops~\cite{Liu2024} exhibit radial and circular cracks); therefore we consider only one particular geometry for the domain. Fractures of brittle materials, which occur due to an impact (glass) or release of residual stress (ceramic glazes), are outside the scope of this work. To reproduce hierarchical crack patterns, three known approaches were used to generate artificial networks: random homogeneous tessellation~\cite{Hafver2014}, recursive Voronoi tessellation~\cite{Tarasevich2025arxiv}, and a simulation model of crack growth~\cite{Yang2025b}. Each algorithm was modified to produce a hierarchy similar to that observed in real systems. Thus, our study extends work~\cite{Roy2022} by investigating a new type of real-world crack patterns and by using a different set of models.

The rest of the paper is organized as follows. Section~\ref{sec:methods} describes the main technical details of real image processing and provides a brief overview of the modeling methods. Section~\ref{sec:results} presents our main findings. Section~\ref{sec:conclusion} summarizes the main results. Appendix~\ref{sec:app} provides a detailed description of all three algorithms used to generate the hierarchical crack networks.

\section{Methods\label{sec:methods}}

\subsection{Sampling and image processing}
For analysis, we choose several images of real-world hierarchical crack patterns, viz., four patterns presented in Ref.~\cite[Fig.~4]{Yang2025} and four patterns presented in Ref.~\cite[Fig.~4]{Yang2025a}. These particular images were chosen because they are of sufficiently high quality for processing and were obtained by the same group using the same methods, which allows all the images to be classified into a single group.
To process these images, we used StructuralGT, a Python-based application for utilizing graph theory analysis to structural networks~\cite{StructuralGT}. StructuralGT converts digital images of structures into a graph theoretical representation, i.e., into sets of nodes and edges.

\subsection{Crack classification}

\begin{figure*}
  \centering
\includegraphics[width=\textwidth]{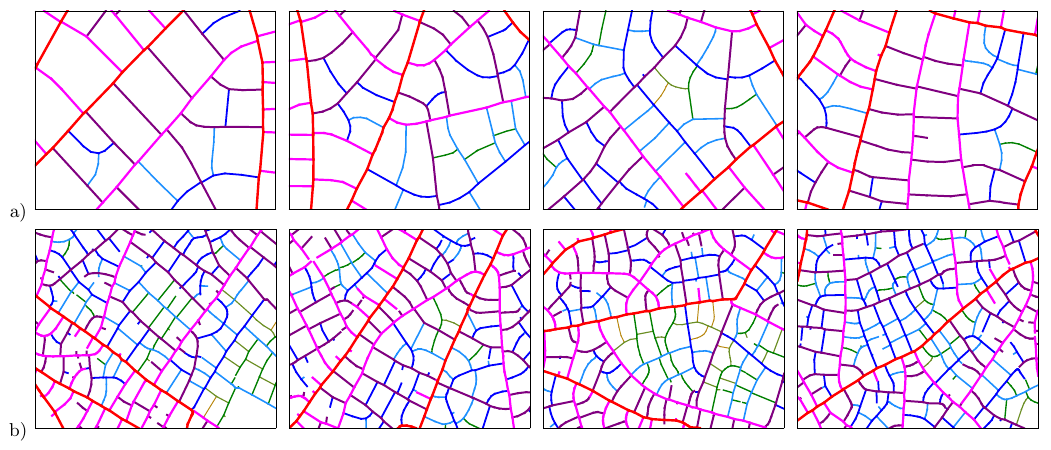}
  \caption{Examples of hierarchical networks obtained from images of real-world crack patterns a) based on~\cite[Fig.~4]{Yang2025}; the sizes of each original image are about $4.4 \times 3.3$~mm; b) based on~\cite[Fig.~4]{Yang2025a}; the sizes of each original image are about $8.9 \times 6.7$~mm. In all cases, classification of crack orders was performed using the algorithm~\cite{Bohn2005}. Cracks of different orders (in ascending order) are painted in the following colors: red, magenta, purple, blue, dodger blue, green, olive drab, and dark goldenrod.\label{fig:HCPYang2025}}
\end{figure*}

Classification of cracks involves two steps. The first step is grouping edges of the graph into chains (segments according to terminology used in~\cite{Perna2011}). When the angle between adjacent edges of a graph is $180^\circ \pm 30^\circ$, they are assumed to be parts of the same chain which corresponds to a crack in the original image of the crack pattern. When all chains are identified, the second step starts. This step was designed to assign an order to each chain according to~\cite{Bohn2005}. To ensure consistency, we treat boundaries of the domain as a crack of zeroth order. When an end of the chain is a dead-end (i.e., degree of the vertex is unity) and another end connects to the chain of order $n$, $(n+1)$-th order is assigned to the former chain. When the chain ends are connected to chains of orders $n$ and $m$, then order $\max(n,m)+1$ is assigned to the chain under consideration. Note that in algorithm~\cite{Perna2011}, the order of each edge is determined at the first step, then the edges are grouped into segments (chains in our terminology), while the order of the edge may change. We should emphasize, that, in fact, orders of cracks are relative since only a small part of the entire crack pattern is presented in an image. In the approach of Ref.~\cite{Perna2011}, a single crack or a set of cracks is chosen as a root, and the orders of the remaining cracks are defined relative to that root. Similarly, we assume that the chain that spans from one boundary of the domain to the other boundary is a crack of order 1 (primary crack). This assumption could be incorrect if the entire crack pattern were available. Examples of the obtained hierarchical networks are shown in Fig.~\ref{fig:HCPYang2025}.

\subsection{Quantitative analysis of the structure of hierarchical networks}
To characterize both the networks corresponding to real-world crack patterns and the computer-generated networks, we used the following quantities: angles between adjacent edges, edge lengths, areas of regions bounded by cracks, the number of vertices in such regions, and the circularity~\cite{Richard2001},  also known as the isoperimetric quotient. The circularity of one particular shape is
\begin{equation}\label{eq:Circularity}
  Q = \frac{4 \pi  A}{ C^2},
\end{equation}
where $C$ is the perimeter, $A$ is the area of the shape. The closer circularity is to 1, the more circular the polygon. This set of characteristics is almost identical to that used in Ref.~\onlinecite{Roy2022}, except that the angular defect was replaced by the angular distribution, which we consider more demonstrative.

\subsection{Algorithms to generate hierarchical crack patterns\label{subsec:algorithms}}

In Ref.~\cite{Hafver2014} an algorithm was proposed to generate crack patterns. The algorithm uses two parameters, variations of which allow generating patterns of different heterogeneity and topology. In our study, we used only one set of parameters, chosen specifically to exclude dead-end cracks and produce homogeneous patterns. For simplicity, we provide in Appendix~\ref{subsec:Hafver} a description of the algorithm for the particular set of parameters we used rather than the original generic model; namely, in the original model, we set $\gamma=1$ (which leads to homogeneous structures) and $\omega=1$ (which prohibits dead-ends). When dead-end cracks are excluded, the algorithm essentially reduces to a random homogeneous tessellation. We hereafter refer to this algorithm as RHT.

The RHT is designed for the generation of hierarchical crack patterns on a square lattice. Initially, there are no cracks, and the sample boundaries serve as the edges. At each step, a new crack nucleates at a random point, with the selection probability proportional to the Euclidean distance to the nearest existing crack or boundary, which indirectly favors larger domains. A random orientation of crack propagation is then chosen, and the crack propagates in both directions from the nucleation point along a straight line until it meets another crack or a boundary; Bresenham's line algorithm is used for drawing. After the crack is added, the distances are recalculated, and the process repeats until the desired number of cracks is reached.

Another algorithm utilizes a concept of Voronoi fractals~\cite{Shirriff1993}. This idea was recently used to simulate sequential fragmentation~\cite{Fanfoni2025}. In developing our own algorithm (see Appendix~\ref{subsec:RVT}), we aimed to ensure that the algorithm reproduces the hierarchical structure of cracks (both temporal and spatial hierarchy). However, we did not aim to reproduce in detail the geometric properties of real crack networks, in which cracks are often curved and meet each other almost perpendicularly. We took as a basis the Voronoi tessellation, which generally correctly reproduces the morphology of crack networks~\cite{Tarasevich2023}, with the exception of the hierarchical structure. Our algorithm imitates the temporal hierarchy of crack formation: a small number of seeds are placed in a given domain, after which Voronoi tessellation is performed (primary cracks). Assuming that each of the cells (subdomains) into which the system is divided becomes independent, the tessellation procedure is repeated for each resulting cell separately until a given number density of cracks is achieved. We hereafter refer to this algorithm as RVT.

The third algorithm in our study is based on an idea, proposed in Ref.~\onlinecite{Yang2025b}. However, we made a number of modifications to the algorithm (see Appendix~\ref{subsec:appYang2025}). Since dead ends in crack networks are highly undesirable for transparent electrodes based on crack patterns, we ensure that a growing crack cannot finish with a dead end. Since we are only interested in hierarchical crack networks, we excluded the possibility of crack growth from nucleation points within the domain; crack growth always begins either from the sample boundary or from another crack. We excluded the interaction of two growing cracks from consideration, since in our approach, a new crack begins to grow only after the previous crack has completely finished its growth. Furthermore, we refined the behavior of a crack as it approaches an existing crack.

The algorithm iteratively divides a domain using randomly growing cracks. A crack starts on a boundary chosen with a probability depending on its length, and grows in steps with small random direction deviations. As it approaches another boundary, the trajectory is adjusted so that the crack terminates perpendicular to it, after which the domain is divided into two independent subdomains. The process continues recursively: a subdomain is selected for division with a probability depending on its area, until specified limits on the minimum subdomain size or the maximum crack order are reached \footnote{See Supplemental Material at [URL will be inserted by publisher] for an animated example of the algorithm in action.}.

The proposed algorithm differs from that described in Ref.~\onlinecite{Yang2025b} in that:
(i) it prohibits crack initiation in internal regions; (ii) it uses a different method for selecting the initiation point of a new crack on a boundary; (iii) it employs a fundamentally different mechanism for hierarchical crack network formation; (iv) it adopts a different termination criterion; (v) it does not model crack width.

The distributions of the quantities under consideration presented in Section~\ref{sec:results} were obtained using 1000 different samples for each of the algorithms RHT and RVT, and 100 different samples for the algorithm based on  Ref.~\onlinecite{Yang2025b}. The means and variances of all characteristics studied are given in the Supplemental Material for each sample, both real-world and computer-generated~\footnote{See Supplemental Material at [URL will be inserted by publisher] for the  means and variances of all characteristics for each sample.}.

\section{Results\label{sec:results}}

\begin{figure*}
  \centering
  \includegraphics[width=\textwidth]{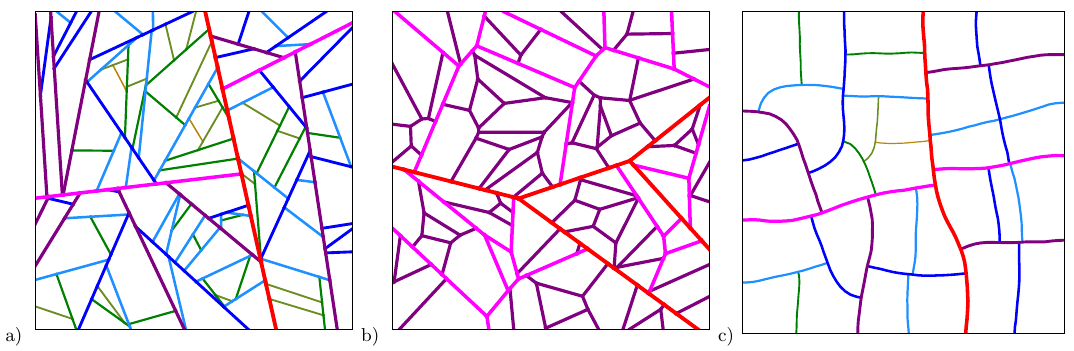}
  \caption{Examples of networks obtained using a) the RHT~\cite{Hafver2014}, b) the RVT, c) our modification of the algorithm~\cite{Yang2025b} . Different crack orders are highlighted using different colors: red, magenta, purple, blue, dodger blue, green, olive drab, and dark goldenrod.}\label{fig:CGPatterns}
\end{figure*}
Figure~\ref{fig:CGPatterns}a) shows an example of a network obtained using our realization of the algorithm~\cite{Hafver2014}.
The edges of the cells are highlighted in different colors according to the classification~\cite{Bohn2005}.
Figure~\ref{fig:CGPatterns}b) shows an example of a network obtained using a RVT as described in Appendix~\ref{subsec:RVT}. The edges of the cells obtained at the first, at the second, and at the third iteration are shown in red, in blue, and in green, respectively. Thus, the coloring of the edges in Fig.~\ref{fig:CGPatterns}b corresponds to the temporal hierarchy, rather than the spatial one, as in Ref.~\onlinecite{Bohn2005}. Figure~\ref{fig:CGPatterns}c) shows an example of a network obtained using the algorithm described in Appendix~\ref{subsec:appYang2025}.

Figure~\ref{fig:CrackAngles} shows distributions of angles between the nearest edges in the networks based on real-world crack patterns along with computer-generated networks. There is a clear similarity in the distributions corresponding to the networks obtained using tessellations (bottom row): a narrow peak at $180^\circ$ and a wide hump centered at $90^\circ$. The visible asymmetry of the hump in the case of networks obtained using the RVT is caused by the presence of Y-type connections, which correspond to angles of about $120^\circ$. There is also a clear similarity between the distributions corresponding to real-world hierarchical crack patterns and the networks obtained using the simulation model: two narrow peaks centered at approximately  $90^\circ$ and $180^\circ$ (upper row). However, there is a significant difference between the distributions of the top and bottom rows. The distributions indicate the presence of $T$-junctions, according to the classification~\cite{Gray1976}. As expected, the tessellations are unable to reproduce correct angle distributions between cracks. However, the angles between cracks are unlikely to have a significant impact on the electrical conductivity of the whole network when its edges are treated as conductors.
\begin{figure*}
  \centering
  \includegraphics[width=\textwidth]{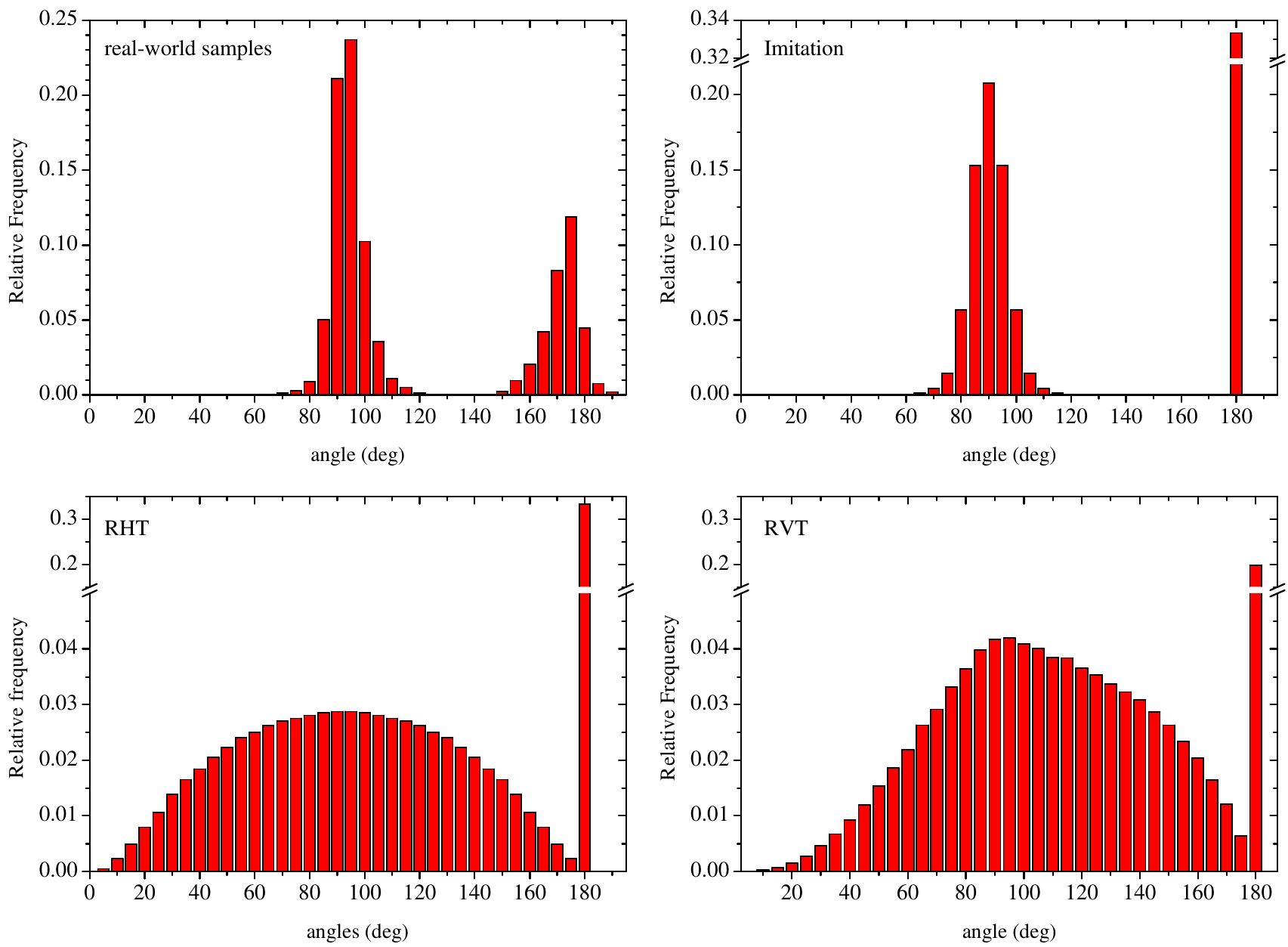}
  \caption{Distributions of angles between cracks. Comparison of results between networks obtained by processing images of real-world crack patterns and computer-generated networks produced by three algorithms.}\label{fig:CrackAngles}
\end{figure*}

\begin{figure*}
  \centering
  \includegraphics[width=\textwidth]{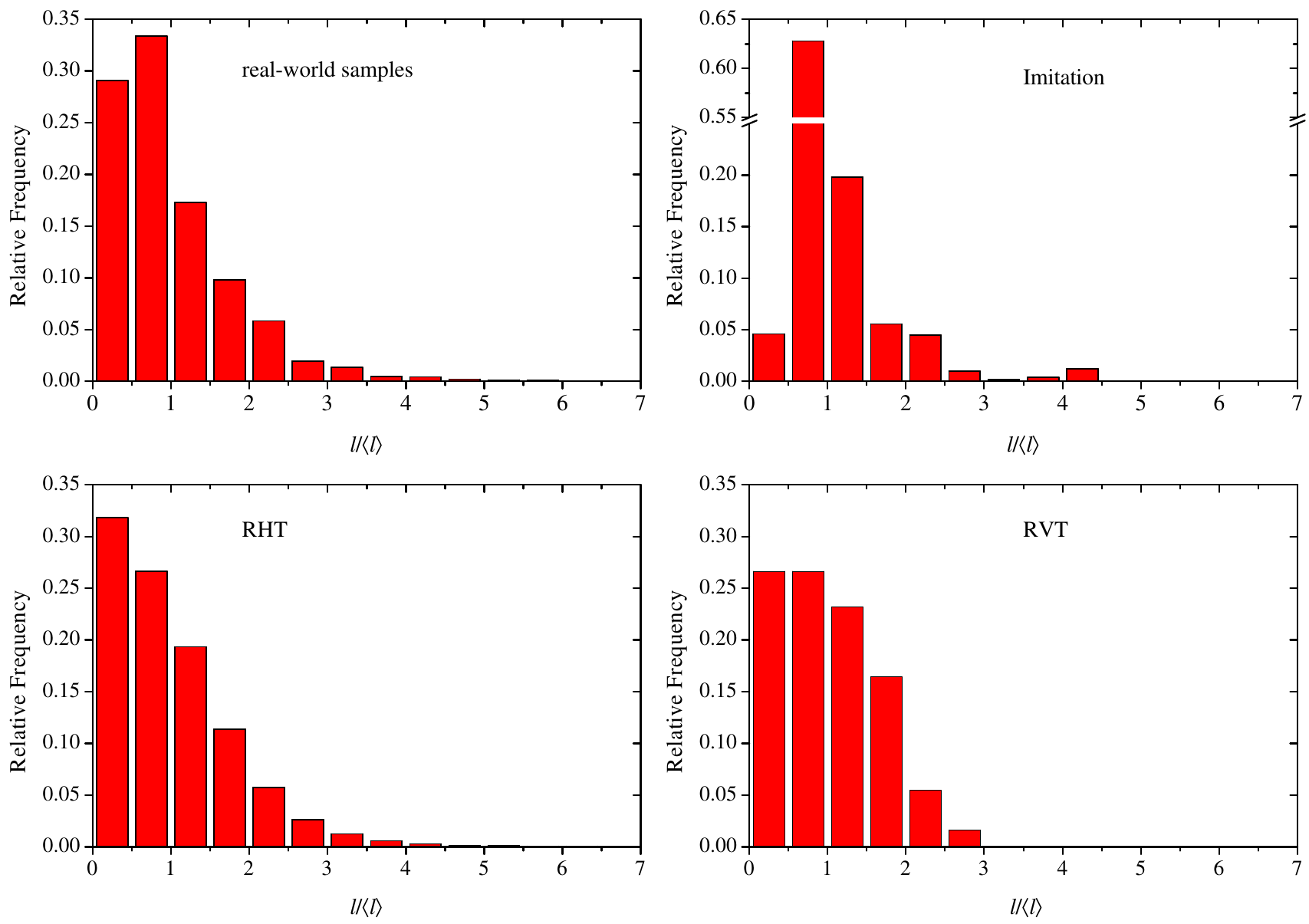}
  \caption{Distributions of edge lengths. Comparison of results between networks obtained by processing images of real-world crack patterns and computer-generated networks produced by three algorithms.}\label{fig:length}
\end{figure*}

Figure~\ref{fig:length} shows distributions of edge lengths. The distributions demonstrate that in the case of the simulation model there are almost no short edges.

\begin{figure*}
  \centering
  \includegraphics[width=\textwidth]{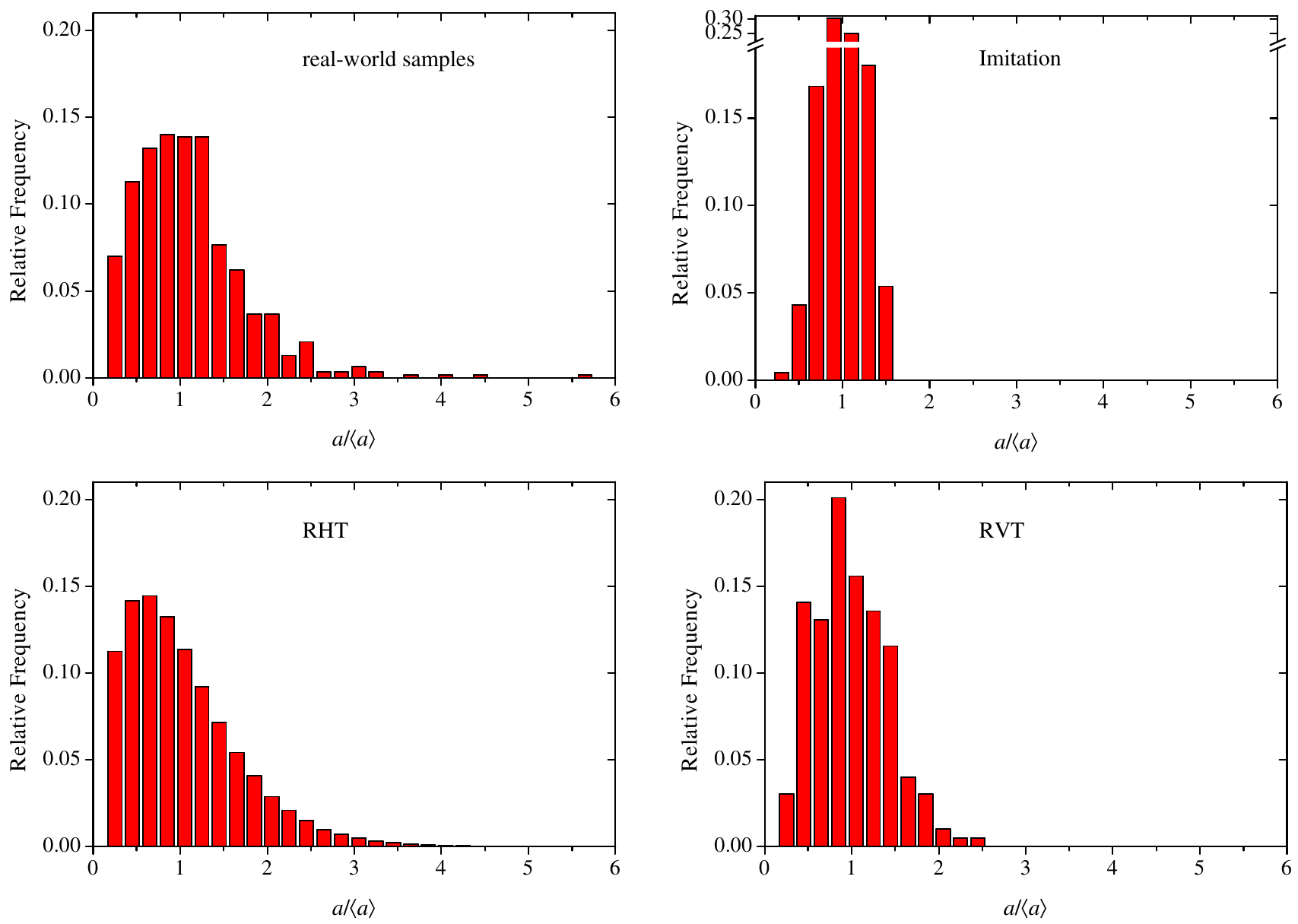}
  \caption{Distributions of cell areas. Comparison of results between networks obtained by processing images of real-world crack patterns and computer-generated networks produced by three algorithms.}\label{fig:areas}
\end{figure*}

Figure~\ref{fig:areas} shows distributions of cell areas. In this case, some similarity between the distributions in the columns is observed. The differences between the distributions in the left column and the right column are quite significant: the distributions in the right column are much narrower and less asymmetrical.

\begin{figure*}
  \centering
  \includegraphics[width=\textwidth]{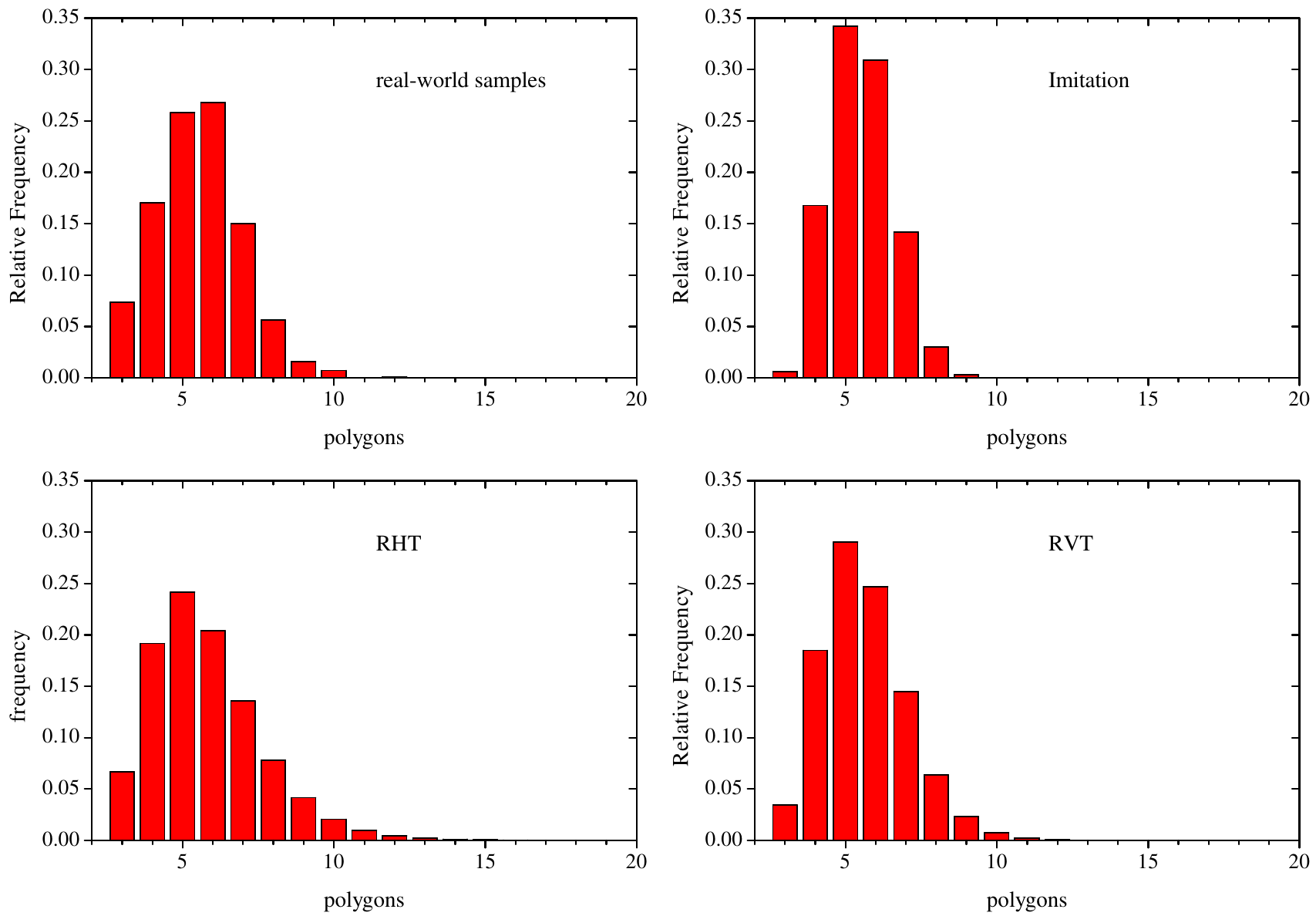}
  \caption{Distributions of cells by number of sides.  Comparison of results between networks obtained by processing images of real-world crack patterns and computer-generated networks produced by three algorithms.}\label{fig:polygons4}
\end{figure*}

Figure~\ref{fig:polygons4} shows distributions of cells by the number of sides. The distribution corresponding to the simulation model is the narrowest, namely, almost all cells have from 4 to 7 sides. The other two models exhibit distributions similar to those in real-world hierarchical crack patterns.

\begin{figure*}
  \centering
  \includegraphics[width=\textwidth]{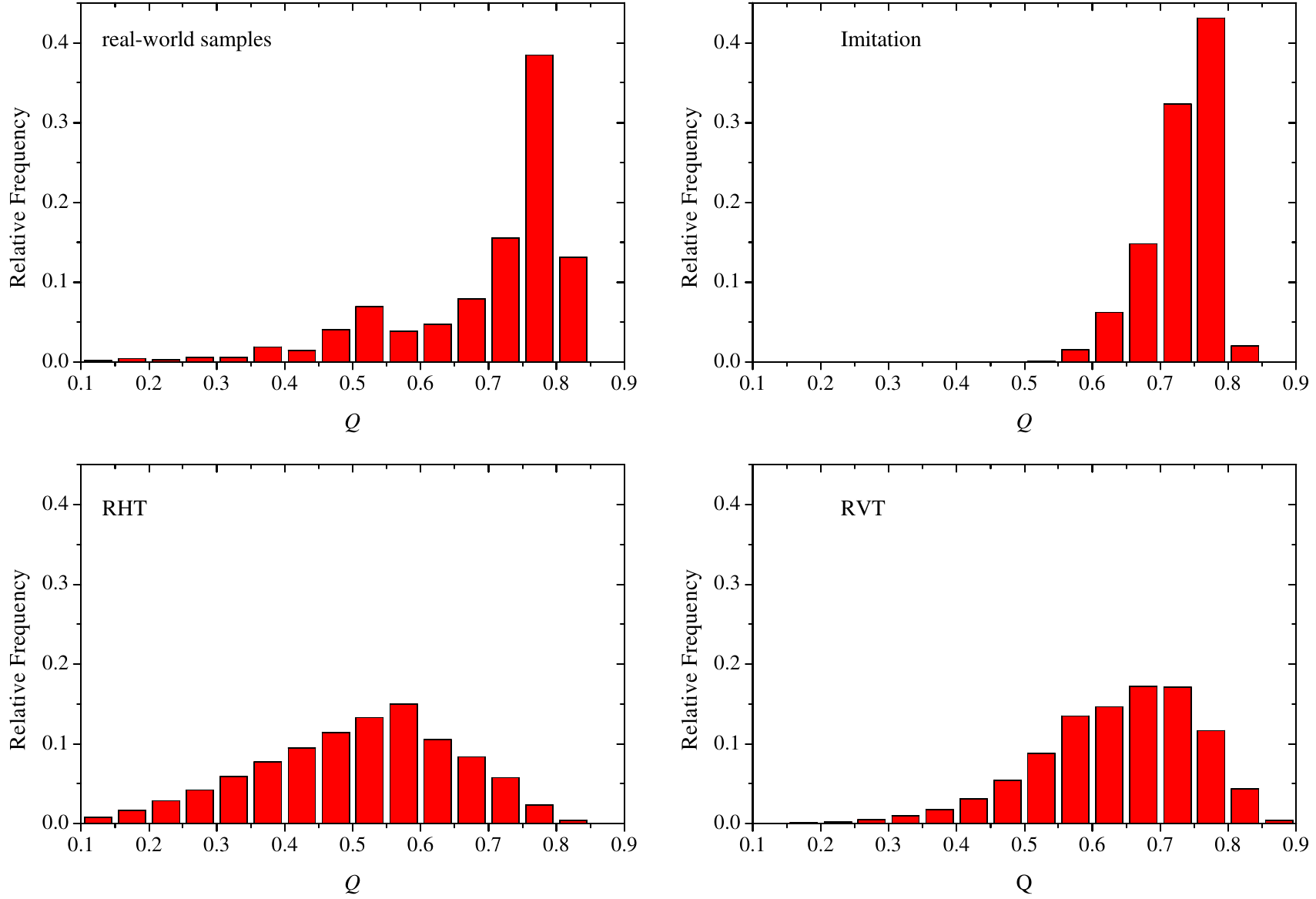}
  \caption{Distributions of cells by the circularity~\eqref{eq:Circularity}. Comparison of results between networks obtained by processing images of real-world crack patterns and computer-generated networks produced by three algorithms.}\label{fig:Shapefactor4}
\end{figure*}

Figure~\ref{fig:Shapefactor4} shows the distributions of cells by the circularity. Networks obtained using the RHT and the RVT (bottom row) exhibit wider and flatter distributions than real-world crack patterns and networks obtained using the simulation model (upper row). However, the distribution obtained from the simulation model is narrower than that obtained from processing real samples.

\section{Conclusion\label{sec:conclusion}}
The study of hierarchical crack patterns forming during the desiccation of thin colloidal or polymer films on solid substrates is important both for understanding the fundamental mechanisms of crack formation and for applied problems in optoelectronics, in particular, for the fabrication of crack-template-based transparent electrodes.

In this work, we analyzed the morphology of real-world hierarchical crack patterns using image processing.
To reproduce hierarchical crack patterns, three known approaches were used to generate artificial networks: RHT~\cite{Hafver2014}, RVT~\cite{Tarasevich2025arxiv}, and a simulation model of crack growth~\cite{Yang2025b}. Each algorithm was modified to produce a hierarchy similar to that observed in real systems. A comparison of the geometric characteristics (distributions of angles between adjacent cracks, edge lengths, cell areas, circularity of cells) and topological properties (distribution of the number of cell sides) of real-world and computer-generated networks reveals that: (i) the simulation model best reproduces the key features of real cracks, including characteristic right angles at intersections and narrow peaks in the distributions; (ii) networks obtained using tessellations yield wider and flatter angle distributions and do not capture the specificity of $T$-junctions; however, their topological characteristics (distribution of the number of cell sides) are closer to real ones than those of the simulation model; (iii) the cell area distributions for all three models are qualitatively similar to the real-world ones, but the simulation model yields the narrowest distribution.

Thus, none of the considered approaches exactly reproduces the entire set of properties of real-world hierarchical crack patterns; however, the simulation model of crack growth demonstrates the best agreement in terms of key geometric features. The obtained results provide a basis for further improvement of modeling methods and the development of automatic classification of hierarchical crack patterns from their images.

While tessellations offer a convenient and simple mathematical model of crack patterns, they inherently fail to capture their detailed properties. A simulation-based approach could potentially reproduce realistic hierarchical patterns, provided its parameters are identifiable. However, this identification requires video recordings of crack development across numerous samples; currently, only images of small fragments of the final patterns are available.

It should be noted that the accuracy of reproducing the properties of real-world hierarchical crack patterns in simulation modeling directly depends on two factors. First, a robust algorithm for crack classification by order is required, ensuring invariance of the results with respect to changes in the size, position, and boundaries of the analyzed image fragment. Second, a representative set of high-quality images of different fragments of the same hierarchical crack pattern is needed. The presence of these two conditions would allow us to identify the simulation model parameters and to obtain computer-generated networks whose geometric and topological properties are expected to be closer to real-world crack patterns than those achieved in the present work.

There is a serious (and in most cases intractable) problem of establishing a one-to-one correspondence between the temporal hierarchy of crack initiation and the spatial hierarchy that can be extracted from an image of a fragment of the entire hierarchical crack pattern. In the case of computer-generated hierarchical crack patterns, the hierarchy of cracks is inherently temporal. As a result, comparing this temporal hierarchy with the spatial hierarchy extracted from images of real specimens becomes extremely difficult. Existing classification methods exhibit a number of serious problems when dealing with images of a fragment of the entire hierarchical fracture network. Without addressing these issues, comparisons cannot be considered reliable and informative.

\begin{acknowledgments}
A.S.B. and A.V.E. acknowledge funding from the Russian Science Foundation, Grant No. 25-21-00460.
Y.Y.T. is thankful to Avik Chatterjee for fruitful discussions at the early stage of this study.
\end{acknowledgments}

\appendix

\section{Algorithms for generating hierarchical crack patterns\label{sec:app}}

The following concepts are used in the descriptions of algorithms.

\emph{Sample} is a part of the plane in which a hierarchical crack network will be generated.

\emph{Domain} is a part of the original sample bounded by cracks or sample boundaries and containing no cracks inside. At the first step of the algorithm, the domain is the entire sample.

\subsection{Algorithm for generating hierarchical crack patterns using a tessellation\label{subsec:Hafver}}

This algorithm is an adaptation of the algorithm~\cite{Hafver2014}. Both the original algorithm and its adaptation are based on the discrete approach, viz., a square lattice $L \times L$ ($L=256$ in our simulations) is used as a sample.  The use of a discrete space (a square lattice) radically reduces the complexity of the algorithm, since at each step the algorithm requires a complete recalculation of all distances, taking into account the newly formed crack. On the other hand, the discretization step is less than 0.5\% of the linear size of the domain, so the error introduced by discretization can be expected to be insignificant.

Initially there are no cracks in this sample.

Set sample boundaries as first and last rows and columns of the grid.

Set the desired number of cracks ($N=300$  in our simulations).

Set the current number of cracks to zero.

Set values of the matrix $\mathrm{D}$ ($L \times L$) as the Euclidean distance to the nearest crack or sample boundary.

Repeat the following steps until the desired number of crack orders is reached.
\begin{enumerate}
\item With probability $P_{i,j}$, a nucleation site  $(i, j) \in \{1, L\} \times \{1, L\}$ is chosen within the sample.
Here,
\begin{equation}\label{eq:PHafver}
P_{i,j} = \frac{d_{i,j}}{Z},
\end{equation}
where
\begin{equation}\label{eq:Zhafver}
   Z = \sum\limits^L_{i,j=1} d_{i,j},
\end{equation}
$d_{i,j}$ is the element of $\mathrm{D}$.
Note that
$$
\sum\limits^L_{i,j=1}  P_{i,j} = 1.
$$
If nucleation site belongs to an existing crack or to a boundary of the sample, a new site is chosen, since, obviously, $P_{i,j} = 0$ in this case.

Although this rule does not explicitly consider domain sizes, it indirectly favors larger domains.
\item A random angle $\theta \in \left[-\frac{\pi}{2};\frac{\pi}{2}\right]$ is chosen with equal probability. This angle sets the orientation of the new crack.
\item From the nucleation site, a line (corresponding to the new crack) is drawn with a given orientation in both directions until each end meets an existing crack or a sample boundary. Using Bresenham's line algorithm, points corresponding this line are marked as the new crack.
    Move step-by-step along the rasterized line and check whether the current point belongs to a set of cracks or boundaries: stop if true.
\item Increase the current number of cracks by one and recalculate all shortest distances $\mathrm{D}$.
\end{enumerate}
\begin{widetext}
\begin{algorithmic}[1]
\Require Lattice size $L$, desired number of cracks $N$
\Ensure Distance matrix $\mathrm{D} \in \mathbb{R}^{L \times L}$ where zero values indicate cracks or sample boundaries
\State Define sample boundaries: first and last rows ($i=1$, $i=L$), first and last columns ($j=1$, $j=L$)
\State $\mathit{current\_cracks} \gets 0$
\While{$\mathit{current\_cracks} < N$}
    \State Compute matrix $\mathrm{D}$ where $d_{i,j}$ is the Euclidean distance from point $(i,j)$ to the nearest point that is either a sample boundary or an already existing crack (i.e., any point with current $d=0$).
    \LComment{On the first iteration, only boundaries exist}
    \State Compute $Z$ according to \eqref{eq:Zhafver}
    \State Choose a random point $(i,j)$ with probability $P_{i,j}$ given by \eqref{eq:PHafver} \LComment{Points with $d_{i,j}=0$ have zero probability and are never chosen}
    \State Choose a random angle $\theta$ uniformly from $[-\pi/2; \pi/2]$
    \State Mark the nucleation site as a crack: set $d_{i,j} \gets 0$ \Comment{Center of the new crack}
    \State Generate a rasterized line from $(i,j)$ in direction $\theta$ in both directions using Bresenham's algorithm.
    \For{each direction (positive and negative)}
        \State Let $(x,y)$ be the next point from $(i,j)$ along the line in this direction (adjacent point)
        \While{$(x,y)$ is inside the sample (not a boundary) and $d_{x,y} > 0$}
            \State Set $d_{x,y} \gets 0$ \Comment{Mark as crack}
            \State Update $(x,y)$ to the next point on the line in this direction
        \EndWhile
        \LComment{Stop when a boundary or an existing crack ($d=0$) is encountered; that point is not marked}
    \EndFor
    \State $\mathit{current\_cracks} \gets \mathit{current\_cracks} + 1$
\EndWhile
\State \Return $\mathrm{D}$
\end{algorithmic}
\end{widetext}

\subsection{Algorithm for generating hierarchical crack patterns using a recursive Voronoi tessellation  (Voronoi fractals)\label{subsec:RVT}}

This algorithm is based on our independent realisation~\cite{Tarasevich2025arxiv} of the RVT. Similar although not identical approach was proposed to modelling a sequential fragmentation~\cite{Fanfoni2025}.

The main idea behind the method is a sequential application of Voronoi tessellation. Initially, several seeds are randomly placed into the sample. These seeds may be produced using either Poisson process or another method to ensure more homogeneous distribution~\cite{Tarasevich2025}. Using these seeds, the standard Voronoi tessellation function divides the sample into several domains. New seeds are added to each domain and Voronoi tessellation is repeated.

The algorithm builds a hierarchy of nested Voronoi diagrams inside a given polygonal domain. All seeds are generated in advance and split into sequential groups (chunks).

First, a Voronoi diagram is constructed using only the initial group of seeds. Since standard Voronoi routines operate on the entire plane, some domains may extending beyond the sample. To confine the diagram to the sample, a domain is clipped if it intersects sample boundaries.

Next, for every domain, the algorithm examines the following group of seeds: those that lie inside the domain are selected. Together with the original seed that generated the domain, they form the input for a new Voronoi diagram, which is again computed on the whole plane and then clipped to the domain's boundaries.

This process repeats for all subsequent seed groups. At each level, only the seeds from the current chunk that fall into a given domain are used, together with the domain's own center. The recursion continues until all chunks have been processed, producing a multi-level tessellation where each domain may contain a finer subdivision.

In our simulation, the initial number of seeds was 4. Following each tessellation, the number of seeds increased by a factor of 4; a total of three tessellations were performed.

\begin{widetext}
\begin{algorithmic}[1]
\Ensure function gen(N)
\LComment{returns a list of N points distributed in the sample}
\Ensure function voronoi(points, outerVertices)
\LComment{performs Voronoi tessellation for given seeds within the polygon outerVertices; returns a structure with field cells, each cell having vertices and center}
\Ensure function PointsInCell(points, cellVertices)
\LComment{returns the subset of points that lie inside the polygon defined by cellVertices}
\State outerVertices $\gets$ vertices of the sample
\State totalSeeds $\gets$ 200 \Comment{example}
\State generations $\gets$ 3 \Comment{number of steps}
\State chunkSizes $\gets$ [10, 40, 150] \Comment{sum = totalSeeds}
\State allSeeds $\gets$ \Call{gen}{totalSeeds}
\State chunks $\gets$ empty list
\State index $\gets$ 1
\For{size $\in$ chunkSizes}
    \State chunks.append(allSeeds[index:index+size-1])
    \State index $\gets$ index + size
\EndFor
\Function{RecursiveVoronoi}{step, points, outerVertices}
    \State vor $\gets$ \Call{voronoi}{points, outerVertices}
    \If{step = generations}
        \State \Return \Comment{base case: no further refinement}
    \EndIf
    \State nextStep $\gets$ step + 1
    \State newPointsFromChunk $\gets$ chunks[nextStep]
    \For{cell $\in$ vor.cells}
        \State center $\gets$ cell.center
        \State newLocalPoints $\gets$ \Call{PointsInCell}{newPointsFromChunk, cell.vertices}
        \State nextPoints $\gets$ [center] $\cup$ newLocalPoints
        \State \Call{RecursiveVoronoi}{nextStep, nextPoints, cell.vertices}
    \EndFor
    \State \Return
\EndFunction
\State \Call{RecursiveVoronoi}{1, chunks[1], outerVertices}
\end{algorithmic}
\end{widetext}

\subsection{Algorithm for generating hierarchical crack networks}\label{subsec:appYang2025}
The proposed algorithm uses some ideas of the algorithm described in the article~\cite{Yang2025b}, so it can be considered as a modification of it, although it has several fundamental differences.

\paragraph{Algorithm prerequisites.}
 The sample is supposed to be simply connected, i.e., contain no holes. In simulations, samples are usually rectangular, but other shapes are possible. In the case of circular samples, the algorithm in its current form cannot generate plausible hierarchical crack networks.

The algorithm is based on the following observations.
\begin{enumerate}
  \item Cracks primarily arise on the boundaries of a domain, although in the real world there are cases where cracks start growing from internal points of the domain (nucleation points). The latter case is not typical for the hierarchical networks of primary interest to us, so we will not consider the possibility of crack nucleation at internal points of the domain.
  \item Crack nucleation is most likely near the middle of a boundary.
  \item Cracks originate from initial nucleation points. A crack grows almost perpendicular to the boundary. During film cracking, the medium is not perfectly homogeneous, and inhomogeneities influence the crack growth direction, but crack growth is inertial. In the drying thin colloidal and polymer films of interest, meandering cracks are atypical, although cracks may bend.
\end{enumerate}

\paragraph{Algorithm.}
\begin{enumerate}
\item
\emph{Selection of the crack nucleation point.} The point from which crack growth begins is located on one of the boundaries of the domain. This boundary is chosen randomly.

The probability of choosing a domain boundary from which crack growth starts is proportional to the boundary length
\begin{equation}\label{eq:Pside}
P_j^\text{border} = \frac{a^k_j}{C},
\end{equation}
where $a_i$ is the length of the $i$-th boundary of the domain, $C = \sum_i a^k_i$ (for $k=1$ this quantity is the perimeter of the domain), $k$ is a parameter regulating the dependence of the crack nucleation probability on the boundary length: for $k=1$ the probability is proportional to length, for $k>1$ longer boundaries are chosen more often, while they are chosen less often for $k<1$. Obviously, in the case of a square sample when generating the primary crack, the sides are chosen with equal probability.

Crack nucleation is most likely near the middle of the boundary, so the coordinate of the crack nucleation point is determined as follows. Let $l$ be the natural coordinate along a boundary, and let the length of this boundary be $L$; then
\begin{equation}\label{eq:CrackOrigin}
l \sim \mathcal{N}(L/2,\sigma_L^2),
\end{equation}
here $\mathcal{N}(L/2,\sigma_L^2)$ is the normal distribution with mean $L/2$ and variance $\sigma_L^2$. Naturally, hereafter a truncated normal distribution should be used, i.e., discarding points that fall outside the permissible range; in this case, if $l \not\in [0;L]$.

\item
\emph{Crack growth.}  The crack growth occurs in discrete steps of fixed length $L_d$. At each step, the growth direction is adjusted to account for random fluctuations that model local material inhomogeneities.
Let $\theta_n$ be the growth direction at the $n$-th step (angle with the $x$-axis), and $(x_n, y_n)$ be the coordinates of the crack tip after $n$ steps.

Before the first step, the initial direction is set as:
\begin{equation}\label{eq:eq3}
  \theta_0 = \phi_\text{boundary} + \Delta \theta_0,
\end{equation}
where $\phi_\text{boundary}$ is the normal to the boundary from which growth begins, and $\Delta \theta_n$ are independent realizations of a random variable distributed according to a truncated normal law
\begin{equation}\label{eq:DeltaTheta}
\Delta \theta_n \sim \mathcal{N}(0,\sigma_\theta^2),
\end{equation}
with truncation ensuring growth into the interior of the domain (for example, $\Delta\theta_n \in [-\pi/2, \pi/2]$). Thus, each crack segment has length $L_d$, and its direction deviates from the previous one by a random angle $\Delta\theta_n$.

Then, for each step $n = 0,1,2,\dots$, the following is performed:
\begin{align}
\theta_{n+1} &= \theta_{n} + \Delta \theta_n,\label{eq:thetan1}\\
x_{n+1} &= x_n + L_d \cos \theta_{n+1},\label{eq:xn1}\\
y_{n+1} &= y_n + L_d \sin \theta_{n+1}.\label{eq:yn1}
\end{align}

If during growth the crack enters the attraction zone of a boundary (see the next section), the direction is additionally adjusted according to the mechanism described there; in these formulas, the current value of $\Delta\theta_n$ is used as an independent realization at each step.

\item
\emph{Crack behavior near the domain boundary and growth termination.}
During crack growth, the shortest distance $d$ from its tip to the nearest domain boundary (excluding the boundary from which the crack started) is computed. When this distance becomes less than \( D_a \), the crack enters the attraction zone of the boundary and subsequently changes direction at each step to, as much as possible, enter the boundary perpendicularly. Here
\begin{equation}\label{eq:Da}
D_a =  F \sum_i a_i,
\end{equation}
where $F$ is a parameter.

As long as \(L_d < d < D_a\), at each step a segment of length \(L_d\) is added to the crack according to formula~\eqref{eq:thetaattract}. If \(d \leqslant L_d\), a final segment of length \(d\) is added to the crack, and its direction is computed using formula~\eqref{eq:thetaattract} with the current distance $d$ (instead of $L_d$) and the corresponding value of \(\Delta\theta_n\) (Fig.~\ref{fig:attzone}).
\begin{figure}[!htb]
  \centering
  \includegraphics[width=0.6\columnwidth]{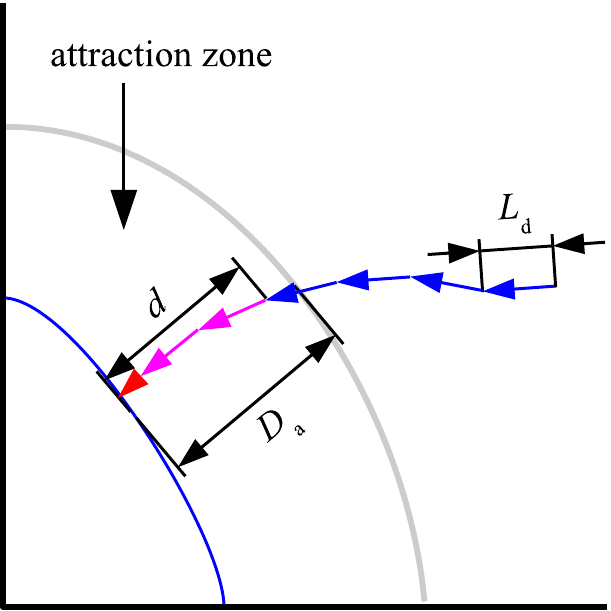}
  \caption{Behavior of a crack in the attraction zone to the boundary of the domain.}\label{fig:attzone}
\end{figure}

Let $\alpha$ be the angle between the normal to the nearest boundary segment from the crack tip and the $x$-axis, and let $\theta_{n}$ be the angle between the current crack growth direction and the $x$-axis (see Fig.~\ref{fig:anglesalgo}). Then $\Delta \alpha = \alpha - \theta_{n}$, and at the next step the crack growth direction is determined by the formula
\begin{equation}\label{eq:thetaattract}
    \theta_{n+1} = \theta_{n} + w \Delta \alpha + \Delta \theta_n,
\end{equation}
where
\begin{equation}\label{eq:weight}
w = 1 - \left( \frac{d}{D_a} \right)^{d_w},
\end{equation}
$\Delta \theta$ is computed according to formula~\eqref{eq:DeltaTheta}, and $d_w$ is a parameter determining the turning speed of the growing crack; the closer the crack tip is to the boundary, the stronger the tendency to move perpendicular to the boundary. During crack development in the attraction zone, the boundary segment nearest to the growing crack tip may change, which does not affect the rules of its growth.
\begin{figure}[!htb]
  \centering
  \includegraphics[width=0.6\columnwidth]{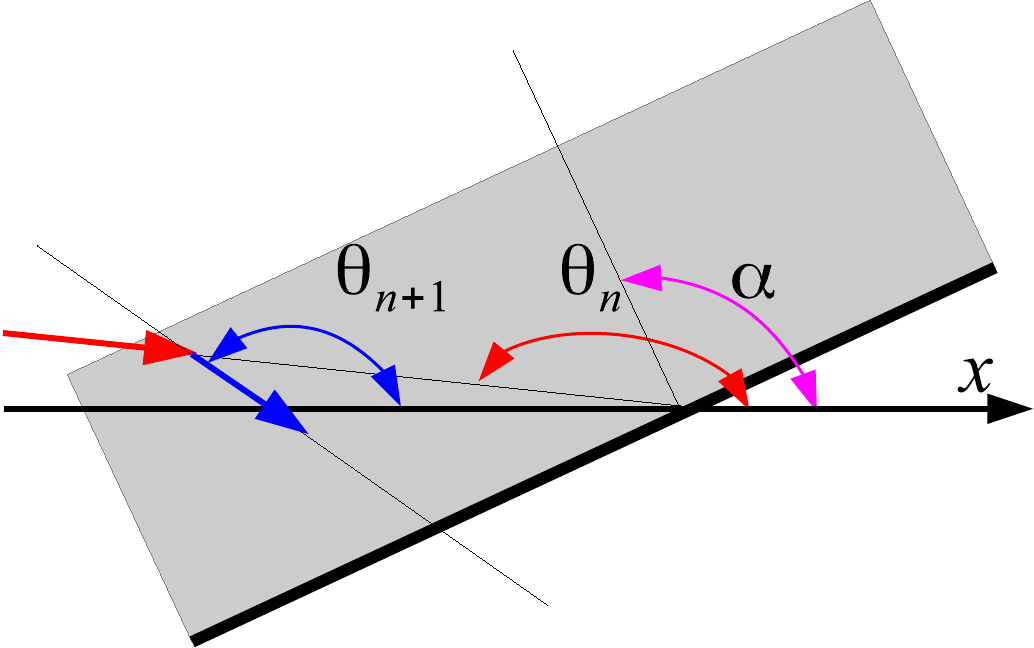}
  \caption{Growth of a crack that has entered the attraction zone.}\label{fig:anglesalgo}
\end{figure}

\item
\emph{Crack growth in domains.}
After the growth of the first crack is completed, the sample is divided into two domains, each of which in no way interacts with the other, and the processes of crack formation in each of the domains are absolutely independent (Fig.~\ref{fig:Step1}). The crack dividing the domains is a first-order crack. It serves as a full-fledged boundary for the two new domains.
\begin{figure}[!htb]
  \centering
  \includegraphics[width=\columnwidth]{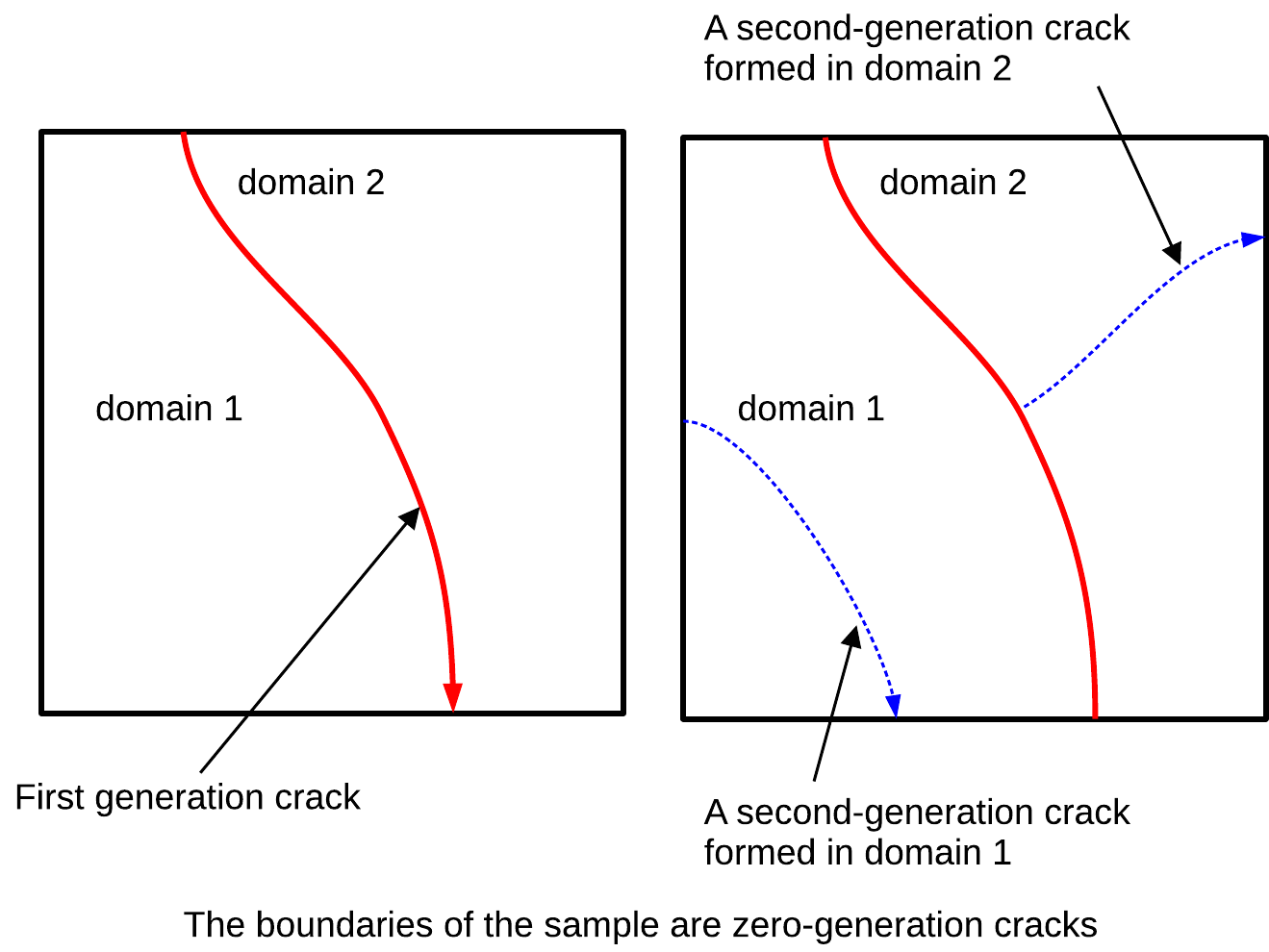}
  \caption{Process of crack network formation.}\label{fig:Step1}
\end{figure}

The initiation and growth of cracks in each domain occur independently. To determine the sequence of domain processing, a domain is chosen with a probability proportional to its area,
\begin{equation}\label{eq:domain}
    P_i^\text{domain} = \frac{\left(S_i^\text{domain}\right)^m}{\sum_j \left(S_j^\text{domain}\right)^m}.
\end{equation}
For $m=1$, the probability of choosing a domain is proportional to its area; for $m>1$, the preference for larger domains increases, and for $m<1$ it decreases.
If a domain was obtained as a result of the appearance of an $n$-th order crack, then the new crack will be of order $n+1$. Within the domain, a boundary from which the growth of a new crack will start is chosen randomly according to formula~\eqref{eq:Pside}. On this boundary, the nucleation point of the new crack is determined by formula~\eqref{eq:CrackOrigin}, and its direction by formula~\eqref{eq:eq3}.

\item
\emph{Termination condition.} If a uniform network is desired, a minimum domain area value $S_\text{min}$ and a limit on the maximum allowable crack order $G_\text{max}$ are specified. The process terminates when the areas of all domains reach a value such that further division would violate the minimum domain size constraint, or when further division of a domain would result in a crack order number exceeding $G_\text{max}$.

If a network with a specified number of crack orders is desired, the process terminates upon reaching the maximum number of crack orders in each domain, i.e., further division would lead to the appearance of cracks of order $G_\text{max} + 1$.
\end{enumerate}

During the simulation, the total length of each crack should be computed; for this purpose, the number of traversed segments of length $L_d$ is stored, and the length of the final crack segment is recorded.

\paragraph{Model parameters \footnote{See Supplemental Material at [URL will be inserted by publisher] for value of each parameter used in our simulation.}:}
\begin{itemize}
\item $L_d$ is crack step size;
\item $F$ is parameter determining the width of the crack attraction zone to the domain boundary;
\item $k$ and $m$ are parameters determining the probability of choosing a boundary and a domain, respectively;
\item $d_w$ is parameter determining the behavior of the crack in the boundary attraction zone;
\item $S_\text{min}$ is minimum allowable domain area;
\item $G_\text{max}$ is maximum allowable crack order;
\item $\sigma_L$ is parameter of the normal distribution, recommended values $\sigma_L \in [L/10;L/6]$;
\item $\sigma_\theta$ is parameter of the normal distribution, recommended values $\sigma_\theta \in [\pi/20;\pi/12]$.
\end{itemize}
In our simulations the values of the parameters were as follows:
$L_d= 0.2$, $F = 0.1$, $k = 10$, $m = 1$, $d_w = 0.5$, $S_\text{min} = 2.5$, $G_\text{max} = 30$, $\sigma_L = 0.03$, $\sigma_{\theta} = 5.0^\circ$.

\cleardoublepage
\begin{widetext}
\begin{algorithmic}[1]
\Require Sample domain, parameters $L_d$, $F$, $k$, $m$, $d_w$, $S_{\min}$, $G_{\max}$, $\sigma_L$, $\sigma_\theta$
\Ensure Set of cracks with assigned generations
\State \textbf{Define structure} \textsc{Domain}:
\State \quad $\textit{level}: \mathbb{N}_0$
\State \quad $\textit{area}: \mathbb{R}$
\State \quad $\textit{boundaries}: \text{list}$
\State \textbf{end}
\State \textbf{Define structure} \textsc{Crack}:
\State \quad $\textit{generation}: \mathbb{N}$
\State \quad $\textit{segments}: \text{list}$
\State \textbf{end}
\State $R  \gets \textsc{Domain}(\textit{level}{=}0,\textit{area}{=}S_0,\textit{boundaries}{=}[\textit{outer\_boundary}])$
\State Initialize cracks list cracks ← []
\While{ $\exists r \in \mathcal{R} : r.level < G_{\max}$ \textbf{and}  ( (mode = uniform \textbf{and} $r.area > S_{\min}$ ) \textbf{or} (mode = generations )) }
    \State Select a domain $r$ from $\mathcal{R}$ with probability proportional to $(r.area)^m$ \Comment{Eq.\ \eqref{eq:domain}}
    \State $P \gets \text{perimeter}(r)$
    \State $D_a \gets F P$
    \State $L \gets \text{len}(b)$
    \State Select a boundary $b$ of $r$ with probability proportional to $L^k$ \Comment{Eq.\ \eqref{eq:Pside}}
    \State Choose a starting point $p_0$ on $b$ from a truncated normal distribution $\mathcal{N}(L/2,\sigma_L^2)$ \Comment{Eq.\ \eqref{eq:CrackOrigin}}
    \State Draw initial noise $\Delta\theta \sim \mathcal{N}(0,\sigma_\theta^2)$ (truncated to keep growth inward)
    \State $\theta \gets \phi_{\text{boundary}} + \Delta\theta$ \Comment{Eq.\ \eqref{eq:eq3}}
    \State $p \gets p_0$ \Comment Set initial direction
    \State Initialize crack segments list $segs \gets []$
     \While{True}
        \State Compute shortest distance $d$ from $p$ to any other boundary of $r$ (excluding the starting one)
        \If{$d \leq L_d$}
            \State \textbf{exit loop} (reached attraction zone boundary)
        \EndIf
        \State Draw new noise $\Delta\theta_n \sim \mathcal{N}(0,\sigma_\theta^2)$ (truncated)
        \If{$d < D_a$} \Comment{Eq. \eqref{eq:Da}}
            \State \text{Compute weight} $w \gets  1 - (d/D_a)^{d_w}$
            \State $\theta \gets \theta + w(\alpha - \theta) + \Delta\theta_n$ \Comment{Eq.\ \eqref{eq:thetaattract}}
        \Else
            \State $\theta \gets \theta + \Delta\theta_n$ \Comment{Eq.\ \eqref{eq:thetan1}}
        \EndIf
        \State $p_{\text{new}} \gets p + L_d(\cos\theta,\sin\theta)$
        \State Append segment $(p, p_{\text{new}})$ to $\textit{segs}$
        \State $p \gets p_{\text{new}}$
    \EndWhile
    \LComment{Add final segment to the boundary}
    \State Compute $\theta_{\text{final}}$ using Eq.\ \eqref{eq:thetaattract} with distance $d$ (now $d \le L_d$)
    \State $p_{\text{final}} \gets p + d(\cos\theta_{\text{final}}, \sin\theta_{\text{final}})$
    \State Append segment $(p, p_{\text{final}})$ to $\textit{segs}$
    \State $\textit{gen} \gets r.\textit{level} + 1$
    \State $\textit{crack} \gets \textsc{Crack}(\textit{generation}=\textit{gen},\ \textit{segments}=\textit{segs})$
    \State Append $\textit{crack}$ to $\textit{cracks}$
    \State Split domain $r$ into $r_1, r_2$
    \State $r_1.\textit{level} \gets \textit{gen}$
    \State $r_2.\textit{level} \gets \textit{gen}$
    \State $r_1.\textit{area} \gets$ Compute area of $r_1$
    \State $r_2.\textit{area} \gets$ Compute area of $r_2$
    \State $r_1.\textit{boundaries} \gets \text{Extract boundaries of } r_1$
    \State $r_2.\textit{boundaries} \gets \text{Extract boundaries of } r_2$
    \State $\mathcal{R} \gets (\mathcal{R} \setminus \{r\}) \cup \{r_1, r_2\}$
\EndWhile
 \State \Return $\textit{cracks}$
\end{algorithmic}
\end{widetext}

\bibliography{cracks}
\end{document}